\documentclass[aps,prb,floatfix,
 amsmath,amssymb,
 reprint,superscriptaddress,
]{revtex4-2}

\usepackage{graphicx}
\usepackage{dcolumn}
\usepackage{bm}
\usepackage{textgreek}
\usepackage[utf8]{inputenc}
\usepackage[LGR,T1]{fontenc}
\usepackage{etoolbox}
\usepackage{float}
\usepackage{hyperref}
\usepackage{textcase}
\usepackage{tikz}
\usepackage{xcolor}
\usepackage{graphicx}
\usepackage[caption=false]{subfig}
\hypersetup{colorlinks=true,linkcolor=blue,filecolor=gray,urlcolor=blue,citecolor=blue}

\DeclareUnicodeCharacter{2212}{\textendash}

\usetikzlibrary{positioning,arrows.meta,fit}
\definecolor{moduleblue}{RGB}{224,236,248}
\definecolor{datagreen}{RGB}{232,245,233}
\definecolor{inputorange}{RGB}{255,243,224}
\definecolor{outputred}{RGB}{252,228,236}
\tikzset{
    fn/.style={draw, rounded corners=3pt, fill=moduleblue, minimum width=3.0cm, minimum height=0.95cm, align=center, font=\sffamily\small},
    data/.style={draw, rounded corners=3pt, fill=datagreen, minimum width=3.0cm, minimum height=0.95cm, align=center},
    bundle/.style={draw, rounded corners=3pt, fill=inputorange, inner xsep=9pt, inner ysep=7pt, align=center},
    output/.style={draw, rounded corners=3pt, fill=outputred, very thick, minimum width=3.0cm, minimum height=1.0cm, align=center},
    arrow/.style={-{Latex[length=2.4mm]}, thick}
}

\begin{document}

\title{Acceleration of CTQMC via Denoising Diffusion Models with Auxiliary Spectral Functions}
\author{Rishi Rao}
\affiliation{ 
Department of Physics, Rutgers University, Newark, New Jersey 07102, USA
}
\affiliation{Condensed Matter Physics and Materials Science Department, Brookhaven National Laboratory, Upton, NY 11973, USA}

\author{Li Zhu}
\email{li.zhu@rutgers.edu}
\affiliation{ 
Department of Physics, Rutgers University, Newark, New Jersey 07102, USA
}%

\date{\today}

\begin{abstract}
Continuous-time Quantum Monte Carlo (CTQMC) impurity solvers are a numerically exact, but expensive, option for solving the Dynamical Mean Field Theory (DMFT) equations, with statistical noise from undersampling presenting a major challenge to low temperature studies. This noise is especially egregious within the sampled self-energy, which can destabilize the self-consistency loop upon insertion into the Dyson equation. In this work, we introduce a neural denoiser for fermionic imaginary time Green's functions based on the diffusion model architecture trained on synthetic data. The network predicts positive and normalized auxiliary spectral functions consistent with the noisy imaginary axis CTQMC data, which is then refined via autodifferentiation to match the measured self-energy, its first two high-frequency moments, and impurity occupation similar to the Constrained Residual Minimization (CRM) method. This leads to Green's functions and self-energies that are causal by design. Through this denoising framework we are able to reduce the required CTQMC measurement steps by a factor of 10-100. We benchmark our method with a half-filled Hubbard model on a Bethe lattice, a two-patch Dynamical Cluster Approximation (DCA) of the two-dimensional Hubbard model, and the SrVO$_3$ system. We reproduce the Mott transition upon increasing Coulomb interaction for the half-filled Hubbard model with only 5000 CTQMC measurements, the pseudogap transition for the DCA two-dimensional Hubbard model with only 5$\times$10$^5$ measurements, and match the spectral function of SrVO$_3$ with 50 times fewer measurements than the reference. Since we generate auxiliary spectral functions, we recover the approximate spectral functions of the systems for free without performing analytic continuation. Analytic continuation is also made significantly more stable through this approach since it ensures the causality of the self-energy. 
\end{abstract}
    
\maketitle

\section{Introduction}
Dynamical mean-field theory (DMFT) and its combination with density functional theory (DFT+DMFT) provide a powerful route to study the electronic structure of correlated materials~\cite{dmftGeorges1996,dmftKotliar2006,ctqmcAnisimov1997}. By reducing the lattice problem from DFT to that of a quantum impurity embedded in a non-interacting bath, realistic multi-orbital systems can be modeled and studied fully $\textit{ab-initio}$. The hybridization expansion version of the continuous-time quantum Monte Carlo (CTQMC) algorithm has been the solver of choice as it yields a technically numerically exact solution to the quantum impurity problem in the limit of infinite sampling~\cite{ctqmcWerner2006,ctqmcGull2011}. Unfortunately, the cost of running CTQMC grows rapidly with the inverse temperature $\beta$ and number of orbitals~\cite{ctqmcGull2007}. Low-temperature phase diagram exploration, large multi-orbital systems, and transitions across phase boundaries remain very expensive to converge due to the large number of CTQMC measurements required. 

A critical source of noise within the DMFT loop is the self-energy generated from the Dyson equation $\Sigma = G_0^{-1} - G^{-1}$. Since the measured Matsubara Green's function itself carries statistical uncertainty, the Dyson equation acts to amplify it at intermediate and high Matsubara frequencies~\cite{improvedHafermann2012,sieKaufmann2019}. With the noisy self-energy used to construct the lattice Green's function for the next DMFT iteration, convergence may be inhibited if sampling is too low. Physically interesting regimes such first-order phase transitions or low temperatures, where autocorrelation times are long, can put strain on computing resources. Current strategies to mitigate this noise, such as tail fitting, improved estimators~\cite{improvedHafermann2012,sieKaufmann2019} and constrained residual minimization (CRM)~\cite{crmLabollita2025} all come with drawbacks. Tail fitting requires knowledge of where the measured self-energy will be within the high-frequency regime, improved estimators come with additional computational cost for sampling, and CRM may take the self-energy out of the causal plane. Noise also presents a major challenge to recovering real-frequency quantities. With a noisy self-energy, analytic continuation back to real frequency becomes an extremely ill-defined problem. 

One strategy to reduce noise is to directly project the imaginary time Green's function back onto the causal space. Representing the imaginary time Green's function using compact bases such as Legendre polynomials~\cite{legendreBoehnke2011}, the intermediate representation~\cite{irShinaoka2017,sparseLi2020}, and the discrete Lehmann representation~\cite{dlrKaye2022} can help preserve most of the elements of causality, but noise reduction by truncation can still break causality due to finite basis error. The Hankel projection method~\cite{hankelYu2024,hankelKemper2024} is another novel idea to project the Green's function to the causal space by finding the closest positive definite Hankel matrix consistent with the data. This has the downside of restricting the resulting imaginary time Green's function to be derived from the closest Hankel matrix, rather than projecting to the space of likely Green's functions, and discarding information on higher-order Matsubara Green's function and self-energy moments. Sparse modeling~\cite{spmOtsuki2017} instead fits a non-negative spectral function to the noisy data, and causal continuation methods such as Nevanlinna interpolation~\cite{nevanlinnaFei2021}, projection-estimation-semidefinite relaxation~\cite{pesHuang2023}, and minimal pole representations~\cite{mprZhang2024} recover a causal Green's function together with its spectrum. While each method has strengths for performing analytic continuation, extremely noisy input data still poses a significant challenge and ensuring consistency with measured data often involves trial and error. On the other hand, neural networks trained on synthetic spectra have been used for analytic continuation for nearly a decade~\cite{acArsenault2017,acYoon2018,acFournier2020,acKliczkowski2024}, with various generation strategies to ensure complete coverage of possible spectral functions.  Although they face various challenges such as sensitivity to the training distribution~\cite{acZhang2022}, difficulties in reproducing Monte Carlo noise~\cite{acSunWang2023}, and loss of physical features~\cite{acAssaad2026}, neural surrogates that replace the impurity solver altogether have reached quantitative accuracy for model and material systems~\cite{mlArsenault2014,mlSturm2021,mlSheridan2021,scalinnLee2025,nnsolverValenti2026}. Warm-start strategies, including our own structure-to-self-energy network~\cite{sigmlRao2025} and intelligent solver initialization~\cite{nainSurrogate2026,makareszRisb2026} reduce the number of DMFT iterations rather than the cost the underlying algorithm. Diffusion models~\cite{ddpmHo2020} in particular were recently used for analytic continuation of bosonic imaginary-time correlation functions \cite{diffusionMeir2026}, showing promise due to their ability to project into the probability space of viable spectra. The approach we take within this study is to synthesize all the information gained within the CTQMC loop to denoise the imaginary-time Green's function using denoising diffusion models. The imaginary time Green's function is represented by an auxiliary spectral function so that it is causal by construction. This scheme has the benefit of both producing a spectral function which serves as a rough estimate of the true spectral function, as well as projecting the Green's function into the causal space. We go further to ensure consistency with the data by running a CRM-style optimization of the auxiliary spectral function against the measured self-energy, its first two high-frequency moments, and impurity occupation. From this we produce not only noiseless imaginary time Green's functions, but also noiseless self-energies that can be used to stabilize the DMFT loop. 

We benchmark our method against 3 relevant systems at low temperature, a half-filled Hubbard model on a Bethe lattice near its Mott insulating transition, a two-patch dynamical cluster approximation (DCA) of the two-dimensional Hubbard model across the pseudogap transition, and a realistic DMFT calculation of the spectral function for SrVO$_3$. In all 3 cases we find significant reduction in the number of Monte Carlo steps required to obtain usable data, on the order of 10-100 times reduction. This substantially reduces the computational burden of running the CTQMC impurity solver while providing noiseless data.  

\section{Method}

\subsection{Overview}

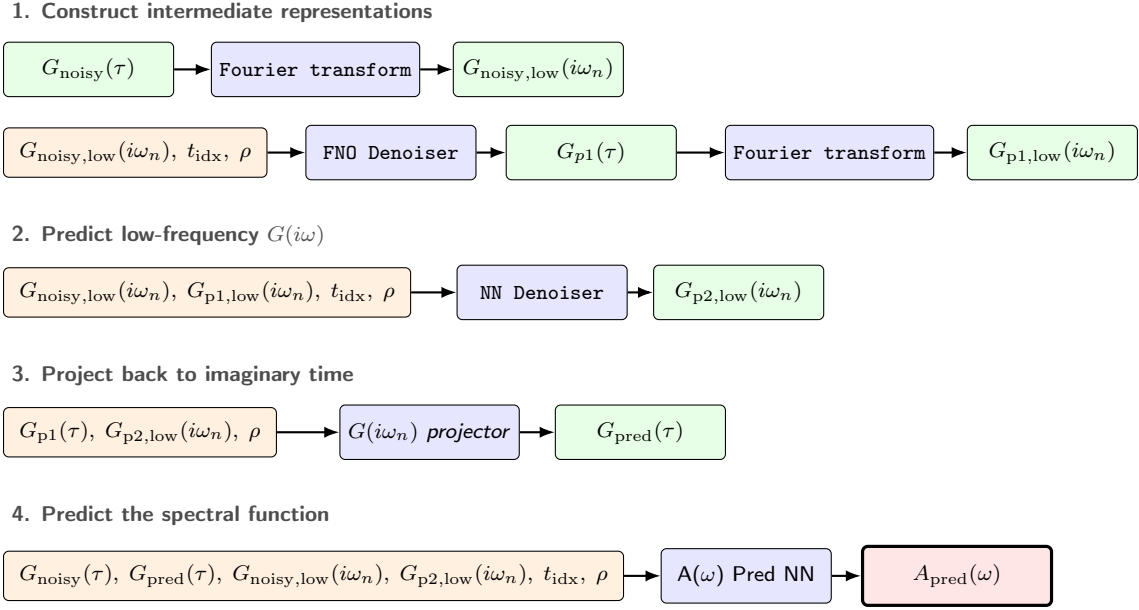
\begin{figure*}[t]
\centering
\begin{tikzpicture}[x=1cm,y=1cm,fn/.style={draw,rounded corners=2pt,fill=blue!10,minimum width=2.25cm,minimum height=0.72cm,align=center,font=\sffamily\footnotesize},data/.style={draw,rounded corners=2pt,fill=green!10,minimum width=2.25cm,minimum height=0.72cm,align=center,font=\footnotesize},bundle/.style={draw,rounded corners=2pt,fill=orange!12,inner xsep=6pt,inner ysep=5pt,align=center,font=\footnotesize},output/.style={draw,rounded corners=2pt,fill=red!10,very thick,minimum width=2.5cm,minimum height=0.78cm,align=center,font=\footnotesize},stage/.style={font=\sffamily\bfseries\footnotesize,text=black!70,anchor=west},arrow/.style={-{Latex[length=2.0mm]},thick}]

\node[stage] at (0,0) {1. Construct intermediate representations};
\node[data,anchor=west] (xt) at (0,-0.75) {$G_{\mathrm{noisy}}(\tau)$};
\node[fn,anchor=west] (low1) at (2.75,-0.75) {\texttt{Fourier transform}};
\node[data,anchor=west] (noisy) at (5.95,-0.75) {$G_{\mathrm{noisy,low}}({i\omega_n)}$};
\draw[arrow] (xt.east) -- (low1.west);
\draw[arrow] (low1.east) -- (noisy.west);
\node[bundle,anchor=west] (basein) at (0,-1.85) {$G_{\mathrm{noisy,low}}({i\omega_n)},\;t_{\mathrm{idx}},\;\rho$};
\node[fn,anchor=west] (base) at (4.00,-1.85) {\texttt{FNO Denoiser}};
\node[data,anchor=west] (rawtau) at (6.65,-1.85) {$G_{p1}(\tau)$};
\node[fn,anchor=west] (low2) at (9.55,-1.85) {\texttt{Fourier transform}};
\node[data,anchor=west] (rawlow) at (12.75,-1.85) {$G_{\mathrm{p1,low}}(i\omega_n)$};
\draw[arrow] (basein.east) -- (base.west);
\draw[arrow] (base.east) -- (rawtau.west);
\draw[arrow] (rawtau.east) -- (low2.west);
\draw[arrow] (low2.east) -- (rawlow.west);

\node[stage] at (0,-2.95) {2. Predict low-frequency $G(i\omega)$};
\node[bundle,anchor=west] (lowheadin) at (0,-3.70) {$G_{\mathrm{noisy,low}}({i\omega_n)},\;G_{\mathrm{p1,low}}(i\omega_n),\;t_{\mathrm{idx}},\;\rho$};
\node[fn,anchor=west] (lowhead) at (6.0,-3.70) {\texttt{NN Denoiser}};
\node[data,anchor=west] (lowpred) at (8.6,-3.70) {$G_{\mathrm{p2,low}}(i\omega_n)$};
\draw[arrow] (lowheadin.east) -- (lowhead.west);
\draw[arrow] (lowhead.east) -- (lowpred.west);

\node[stage] at (0,-4.80) {3. Project back to imaginary time};
\node[bundle,anchor=west] (projin) at (0,-5.55) {$G_{\mathrm{p1}}(\tau),\;G_{\mathrm{p2,low}}(i\omega_n),\;\rho$};
\node[fn,anchor=west] (projector) at (4.45,-5.55) {$G(i\omega_n)$ \textit{projector}};
\node[data,anchor=west] (taupred) at (7.30,-5.55) {$G_{\mathrm{pred}}(\tau)$};
\draw[arrow] (projin.east) -- (projector.west);
\draw[arrow] (projector.east) -- (taupred.west);

\node[stage] at (0,-6.65) {4. Predict the spectral function};
\node[bundle,anchor=west] (awin) at (0,-7.45) {$G_{\mathrm{noisy}}(\tau),\;G_{\mathrm{pred}}(\tau),\;G_{\mathrm{noisy,low}}(i\omega_n),\;G_{\mathrm{p2,low}}(i\omega_n),\;t_{\mathrm{idx}},\;\rho$};
\node[fn,anchor=west] (awhead) at (8.70,-7.45) {A($\omega$) Pred NN};
\node[output,anchor=west] (awpred) at (11.35,-7.45) {$A_{\mathrm{pred}}(\omega)$};
\draw[arrow] (awin.east) -- (awhead.west);
\draw[arrow] (awhead.east) -- (awpred.west);
\end{tikzpicture}

\caption{Computation flow used to obtain the predicted spectral function $A_{\mathrm{pred}}(\omega)$. Input to neural network is noisy imaginary time Green's function $G_{\mathrm{noisy}}(\tau)$, predicted density of impurity orbital $\rho$ and estimated diffusion timestep $t_{\mathrm{idx}}$. The fourier transform step only retains the first $K_{\mathrm{low}}$ Matsubara frequencies. The projector step is a constrained least squares correction ensuring data consistency. } 
\label{fig:aw_pred_nn_diagram}
\end{figure*}

Shown in Figure \ref{fig:aw_pred_nn_diagram} is our flowchart for the transformations performed to eventually arrive at a predicted spectral function. Our multi-stage neural network architecture uses internal representations of the denoised imaginary time and frequency Green's functions to build up enough data to obtain a reasonable estimate of the spectral function. The input of the neural network consists of a noisy imaginary time Green's function $G_{\mathrm{noisy}}(\tau)$, density of impurity orbital $\rho$ and estimated diffusion timestep $t_{\mathrm{idx}}$. $\rho$ can be obtained from the density matrix, which is measured to a much higher accuracy within CTQMC. 

Internally, we predict denoised imaginary time Green's function as well Matsubara frequency Green's function, which is fed into the final $A(\omega)$ prediction network along with the original noisy counterparts, the density and the diffusion timestep to obtain a predicted spectral function. We find it more stable to include a prediction of the first $K_{\mathrm{low}}=16$ positive Matsubara frequencies $G_{\mathrm{p2,low}}(i\omega_n)$ to preserve the overall shape of $G(\tau)$ and the spectral weight near the Fermi level. We include both an internally denoised version $G_{\mathrm{p1,low}}(i\omega_n)$ obtained using a Fourier neural operator (FNO)~\cite{fnoLi2021} and the raw fourier transform $G_{\mathrm{noisy,low}}(i\omega_n)$. The ``NN Denoiser'' step is just a simple feed-forward neural network that combines both the raw Matsubara frequency $G(i\omega_n)$ with the denoised version. The FNO denoising step produces a denoised $G_{p1}(\tau)$ as an intermediary, which is then refined using the denoised low frequency Matsubara Green's function $G_{\mathrm{p2,low}}(i\omega_n)$ through a projection step. Finally, all these intermediaries along with the original input are fed into a fully connected neural network to predict $A(\omega)$. 

\subsection{Training Data}
Similar to previous works on analytic continuation using neural networks~\cite{acArsenault2017,acYoon2018,acFournier2020}, we use sums of Gaussians to generate our synthetic spectral functions. Specifically, we define 
\begin{equation}
    A(\omega) = \sum_{r=1}^{R} a_r \exp\left[{-\frac{1}{2}\left(\frac{\omega-\omega_r}{\sigma_r}\right)^2} \right]
\end{equation}
where $R$ is allowed to vary between 3 and 24, centers are drawn randomly from $|w_r|<8$, the standard deviation $\sigma_r$ is between 0.02 and 1, and $a_r$ is defined such that the spectral functions were positive and normalized. Our grid is hyperbolic with 1001 points defined between $\omega\in\pm$10 and higher density near $\omega=0$. We also explicitly enforce an insulating solution on 15\% of the training data by smoothly reducing the spectral weight to zero for $|\omega|<0.5$. The corresponding imaginary time and frequency Green's functions can easily be obtained through integration. The total training set contains 524,288 spectra, with 32,768 additional spectra generated for the test set. 

\subsection{Diffusion noise schedule and loss}

We train our denoiser by corrupting a clean $G(\tau)$ with curvature-dependent noise, predicting an auxiliary spectral function, and integrating that to obtain the predicted clean $G(\tau)$. $G(\tau)$ is corrupted according to
\begin{equation}
    \bar{G}_t(\tau)=G(\tau)+\sigma_t
\left[
c_0+c_1
\frac{\left|d^2G(\tau)/d\tau^2\right|}
{\max_\tau\left|d^2G(\tau)/d\tau^2\right|}
\right]\epsilon(\tau)
\end{equation}
where $\sigma_t$ controls the amount of noise at a diffusion timestep $t$ and $\epsilon$ is Gaussian noise. We also choose to make our noise curvature dependent as we have found that noise increases near $0$ and $\beta$ where curvature is largest. $c_0$ and $c_1$ are parameters controlling the curvature dependence of the noise. 
Additional complex noise is introduced directly into the low-frequency Matsubara Green's function. The noisy imaginary-time function is first transformed to Matsubara frequency space after which the first $K_{\mathrm{low}}$ Matsubara frequencies are perturbed as
\begin{equation}
    \bar{G}(i\omega_n) + \sigma_{n,t}\xi_n
\end{equation}
where $\xi_{n}$ is complex Gaussian noise and $\sigma_{n,t}$ is scaled according to the average value of the Matsubara Green's function within the training set. This was done to ensure that the high frequency tail, which decays as the inverse of Matsubara frequency, does not experience significantly more noise proportional to its value. The noisy imaginary time Green's function is then projected onto this perturbed $\bar{G}(i\omega_n) + \sigma_{n,t}\xi_n$ such that they are equivalent up to a Fourier transform. 
During the training loop, we sample diffusion timesteps semi-randomly, biased towards larger diffusion timesteps since higher noise is a more challenging regime. 

Our loss function is defined as 
\begin{equation}
\begin{aligned}
\mathcal{L} = 50\mathcal L_{\mathrm{low}}+3\mathcal L_{\mathrm{mid}}+25\mathcal L_{\mathrm{head}}+0.35\mathcal L_{\tau} \\+0.15\mathcal L_{\mathrm{curv}} +2\mathcal L_{\mathrm{end}}+50\mathcal{L_A}
\end{aligned}
\end{equation}
$\mathcal{L}_{\mathrm{low}}$ and $\mathcal{L}_{\mathrm{mid}}$ are the losses for the low ($K_{\mathrm{low}}=16$ and middle ($K_{\mathrm{mid}}=48$) frequency portions of the Matsubara Green's function obtained from Fourier transforming $G_{\mathrm{pred}}(\tau)$. $\mathcal{L}_{\mathrm{head}}$ is the loss for the internal prediction of $G_{\mathrm{p2,low}}(i\omega_n)$. $\mathcal{L}_\tau$ is the smooth L1 loss for the denoised $G_\mathrm{pred}(\tau)$ and $\mathcal{L}_{\mathrm{curv}}$ is the L1 loss for the curvature defined as
\begin{equation}
\mathcal L_{curv} = \frac{1}{N_\tau}\sum_\tau\left| \frac{1}{\sigma_{\mathrm{curv}}(\tau)}\left(
\frac{d^2G_{\mathrm{pred}}(\tau)}{d\tau^2}-
\frac{d^2G(\tau)}{d\tau^2} \right)
\right|,
\end{equation}
where $\sigma_{\mathrm{curv}}(\tau)$ is a curvature scaling parameter. The curvature loss helps to suppress spurious oscillations. $\mathcal{L}_{\mathrm{end}}$ is the loss of $G_\mathrm{pred}(\tau)$ at $\tau=0$ and $\beta$ in order to make sure total density is correct. Finally, $\mathcal{L}_A$ is the loss for the auxiliary spectral function prediction, 
\begin{equation}
\mathcal{L}_A = \frac{1}{N_\omega}\sum_\omega w_A(\omega)\left[\frac{A_{\mathrm{aux}}(\omega)-A(\omega)}{s_A(\omega)}\right]^2
\end{equation}
where the frequency-dependent weight $w_A(\omega)$ is chosen in a way that strongly emphasizes the region near the Fermi level,
\begin{equation}
w_A(\omega)\propto1+50e^{-\frac{1}{2}(4\omega)^2}+8e^{-\frac{1}{2}\omega^2},
\end{equation}
and $s_A(\omega)$ is another scaling parameter. The training therefore preferentially penalizes errors in low-energy features.

The comparatively large weights assigned to $\mathcal L_{\mathrm{low}}$ and $\mathcal L_A$ reflect the goal of preserving low frequency information while accurately reconstructing the true spectral function. By including both terms within our loss, the model can learn analytic continuation along with Green's-function denoising, rather than performing these functions separately. 

Training has been performed using 200 diffusion timesteps over 450 epochs. Training took approximately 30 hours and was performed on a single A100 GPU for an inverse temperature of $\beta=200$. 

\subsection{Noise level estimation}
In general, the noise level of imaginary axis data generated from MC sampling is unknown. This presents a problem as diffusion models require information on the noise level in order to gauge the starting diffusion timestep. We tackle this issue by performing a Hankel projection of the imaginary time Green's function \cite{hankelYu2024} to obtain a baseline $G(\tau)$, and estimating noise as the variance around this as 
\begin{equation}
\sigma_t \approx\frac{\operatorname{median}_{\tau}\left[\left|\dfrac{\left(G(\tau)-G_{\mathrm H}(\tau)\right)-\mathcal{S}\left[G(\tau)-G_{\mathrm H}(\tau)\right]}{\epsilon(\tau)}\right|\right]}{0.67449\sqrt{1-\frac{1}{W}}}
\end{equation}
where $G_{\mathrm H}(\tau)$ is the Hankel-projected Green's function, $\mathcal{S}$ is a moving-average filter of width W, and $\epsilon(\tau)$ is the curvature based noise generated from $G_{\mathrm H}(\tau)$. The factor 0.67449 converts the median absolute deviation to the standard deviation for Gaussian noise, while the factor $\sqrt{1-1/W}$ corrects for subtracting the local moving average. $\sigma_t$ is then matched to the closest diffusion timestep $t$.

\subsection{Refinement of the internal A($\omega$)}
To guarantee that the auxiliary spectral functions are consistent with the CTQMC data, we also perform an optimization against the Dyson equation, as introduced in Ref \cite{crmLabollita2025}. This allows us to include moment information naturally by defining it as an optimization target. 
Our approach differs from Ref \cite{crmLabollita2025} in that we use the internal spectral function representation generated by our diffusion model to obtain $G_{\mathrm{aux}}(i\omega_n)$. Our optimization is then given by 
\begin{equation}
    \begin{aligned}
    \operatorname{min}_{A(\omega)}\left|\mathcal L_{\mathrm{Dyson}}+\mathcal L_{\mathrm{\Sigma_\infty}}+\mathcal L_{\mathrm{\Sigma_1}}+\mathcal L_{\rho}\right|
    \end{aligned}
\end{equation}
where $\mathcal L_{\mathrm{Dyson}}$ is the loss from the Dyson equation, $\mathcal L_{\mathrm{\Sigma_\infty}}+\mathcal{L}_{\mathrm{\Sigma_1}}$ is the loss from the first 2 self energy moments, $\mathcal L_\rho$ is the loss from matching the density, and $A(\omega)$ is treated as a parameter to be optimized.
The Dyson loss term comes from integrating the internal $A_{\mathrm{aux}}(\omega)$ to obtain $G_{\mathrm{aux}}(i\omega_n)$, which is then inserted into the Dyson equation as in Ref \cite{crmLabollita2025}
\begin{equation}
\begin{aligned}
    \mathcal L_{\mathrm{Dyson}} = | G_{\mathrm{aux}}(i\omega_n) - G_0(i\omega_n)\Sigma(i\omega_n)G_{\mathrm{aux}}(i\omega_n) - \\
    G_{0}(i\omega_n) |^2
\end{aligned}
\end{equation}
Where $G_0$ is the Weiss field of the current iteration. The $\mathcal L_{\mathrm{\Sigma_\infty}}+\mathcal{L}_{\mathrm{\Sigma_1}}$ term is obtained from fitting the tail of the Dyson inverse self-energy generated as $\Sigma_{\mathrm{aux}} = G_0^{-1} - G^{-1}$ and calculating the mean absolute squared error. Since self-energy moments can be measured to much higher precision than the actual self-energy during the CTQMC run, this presents an ideal training target to constrain the high frequency behavior. 
The density loss is more direct and can be computed by integrating the spectral function
\begin{equation}
    \mathcal L_{\mathrm{\rho}} = \left| \rho_{CTQMC} - \int_{-\infty}^{\infty} f(\omega) A(\omega) d\omega \right|^2
\end{equation}
where $\rho_{CTQMC}$ is the density produced by measuring the density matrix within CTQMC and $f(\omega)$ is the Fermi-Dirac distribution at the inverse temperature of the calculation. 

We perform all these calculations within PyTorch, which gives us access to numerically exact gradients through autodifferentiation. Since we can backpropagate the gradients all the way to $A_{aux}(\omega)$, we set up an optimization loop to ensure that the internal spectral function representation is consistent with not only the measured self-energy, but also its moments and the measured density. In this manner we can obtain an internal spectral function representation which can be used to generate completely causal self-energies through the Dyson equation without the oscillations that can arise when fitting an unconstrained pole or Lehmann expansion to noisy data \cite{seproofStaar2014}.

\subsection{Impurity Solver}
We have chosen the TRIQS/CTHYB \cite{triqsParcollet2015,cthybSeth2016} solver within the TRIQS package to perform the CTQMC measurements. For each DMFT iteration, we run $N_\mathrm{moves}$ proposals before taking a measurement, meaning the true number of CTQMC steps is actually $N_\mathrm{measure} \times N_\mathrm{moves}$. Our method does not reduce $N_\mathrm{moves}$ as strong auto-correlation can lead to systematic, rather than random, distortion of the Green's function. We also do not consider off-diagonal components of the Green's function, and instead denoise each diagonal component separately in orbital space. 

\begin{figure*}[!t]
    \centering
    \includegraphics[width=2.0\columnwidth]{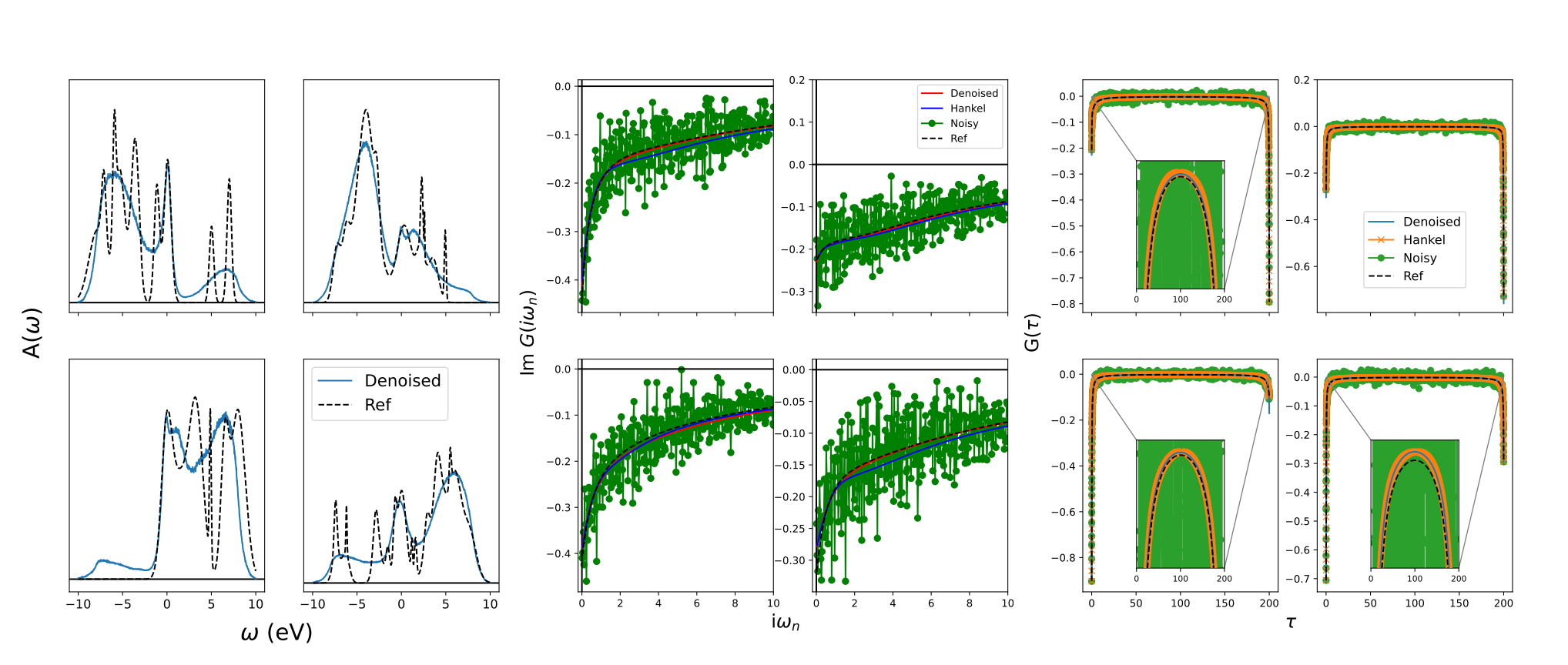}
    \caption{Results of denoising by model without optimization towards self-energy target for 4 randomly chosen examples within a held-out test set.}
    \label{fig:model_eval_results}
\end{figure*}

\begin{figure}[!t]
    \centering
    \includegraphics[width=1.0\columnwidth]{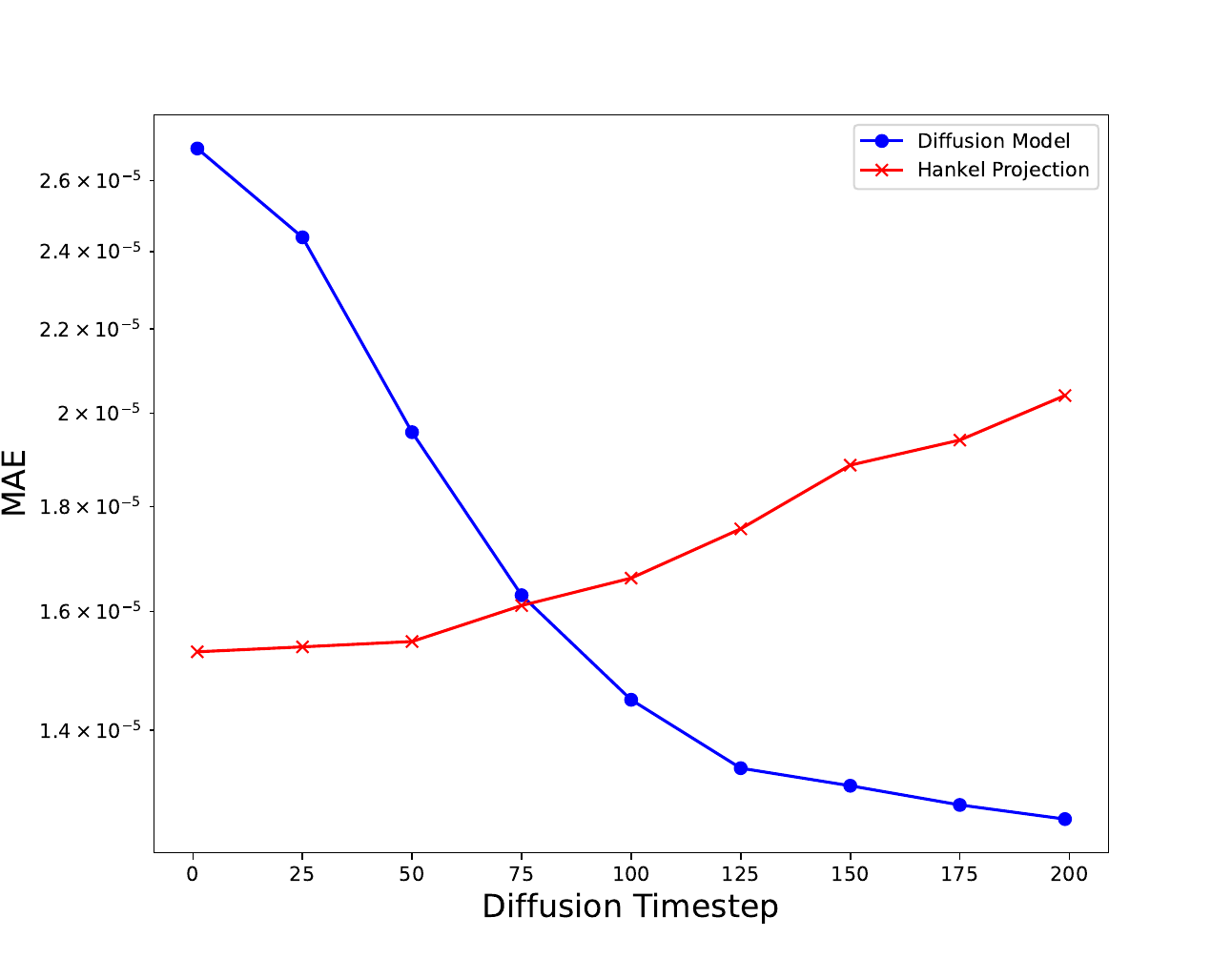}
    \caption{Mean absolute error of denoised $G(\tau)$ versus diffusion timestep for both our denoising model and a Hankel projection. Diffusion timestep here serves as a proxy for noise, with larger diffusion timestep indicating noisier input.}
    \label{fig:errorvdifftime}
\end{figure}

\section{Results}
\subsection{Model Evaluation}
In order to gauge the accuracy of our denoising diffusion model, we benchmark its performance in predicting the true spectral function, $G(\tau)$, and the low frequency portion of $G(i\omega_n)$. Results for 4 randomly chosen samples from the test set at a noise level near the maximum is shown in Fig.~\ref{fig:model_eval_results}. For both $G(\tau)$ and $G(i\omega_n)$, the diffusion model does a better job at matching the original reference function than a standard Hankel projection, with accuracy more apparent for $G(i\omega_n)$. The fine details and peak positions of the spectral function are still unable to be captured by our auxiliary representation, as expected with heavy noise, but the general features are correct near the Fermi level. We plot the mean absolute error of $G(\tau)$ with respect to the diffusion timestep in Fig.~\ref{fig:errorvdifftime} for both our denoising diffusion model and for the Hankel projection. For low noise scenarios, the Hankel projection outperforms the diffusion model, but the diffusion model actually decreases its standard error as timestep (noise) increases. This is likely due to the increased focus on high timestep samples. 

\subsection{Single band Hubbard Model}
We begin our investigation with the Hubbard model at half filling in the paramagnetic phase. We aim to reproduce the insulator to metal transition as a function of local Coulomb interaction U. Our model Hamiltonian is given as 
\begin{equation}
H = -\sum_{<i,j>,\sigma} t_{i,j} c_{i,\sigma}^\dagger c_{j,\sigma} + U \sum_i n_{i\uparrow} n_{i\downarrow}
\end{equation}
which can be solved exactly on a Bethe lattice in the limit of infinite dimensions using DMFT \cite{dmftGeorges1996}. In this limit, the density of states reduces to a semicircular form $\rho(\epsilon) = \frac{1}{2\pi t^2}\sqrt{4t^2 - \epsilon^2}, |\epsilon|<2t$ where we fix $t=0.5$ and define the chemical potential as $\mu = U/2$.

This maps the original Hubbard Hamiltonian to a single impurity Anderson model with a relatively simple hybridization given by $\Delta(i\omega) = t^2G(i\omega)$. We choose $\beta=200$ for our test as lower temperatures generally require more CTQMC steps to converge. This model displays a first-order Mott transition when increasing Coulomb interaction, along with a metal-insulator coexistence region between $U_{c_1} \approx 2.4$ and $U_{c_2} \approx 2.68$ for our chosen value of $\beta$~\cite{ctqmcWerner2006,mitJoo2001,mitRozenberg1999,mitBulla2001,strandDmft2011}. We scan across the phase diagram from $U=2$ to $U=2.8$, reusing the previous hybridization so that the phase within the coexistence regime is metallic. 

\begin{figure*}[!t]
    \includegraphics[width=\textwidth]{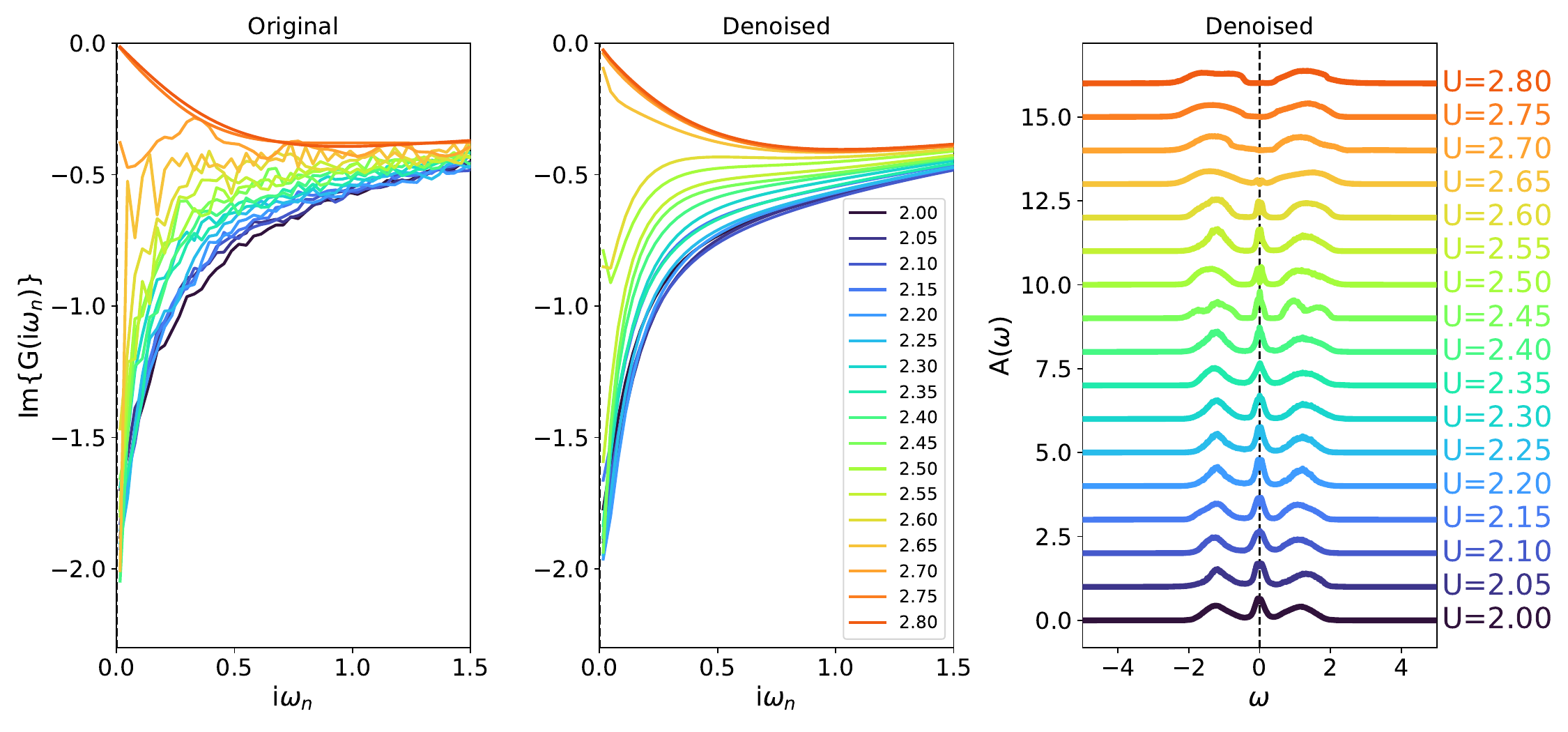}
    \caption{(a) Denoised Green's function versus (b) raw CTQMC Green's function for the last DMFT iteration as a function of Hubbard interaction U for the single band Hubbard model. Also shown in (c) the evolution of the internal spectral function representation for the denoising model. The transition from paramagnetic metal to insulator happens around a U value of 2.65-2.7. Only 5000 CTQMC measurement steps were performed. }
    \label{fig:siamdenoiserawcompare}
    \centering
\end{figure*}

For each DMFT iteration, we test using only 5000 CTQMC measurements with 200 steps between measurements. 10 DMFT iterations are performed for each value of U. This takes a negligible amount of time on modern computing hardware, but generates extremely noisy Green's functions and self-energies as seen in Fig.~\ref{fig:siamdenoiserawcompare}(a). 

With denoising however, we are able to obtain smooth and causal Green's functions as shown in Fig.~\ref{fig:siamdenoiserawcompare}(b). Shown in Figure \ref{fig:siamdenoiserawcompare}(b) and (c) are our results for the Matsubara Green's function and internal $A(\omega)$ when scanning from $U=2.0$ to $U=3.0$. 
When approaching the transition boundary, the noise for the raw CTQMC measurement data goes up significantly, a function of critical slowing down near the Mott transition~\cite{strandDmft2011,zitkoBroyden2009}. In contrast, the denoised results stays smooth and converges rapidly since it is causal by construction and tied to the more accurately measured moments of the self-energy, Green's function and the density. 
The metal to insulator transition is evident not only through the imaginary part of the Green's function, but also through the internal spectral function representation which shows the formation of an insulating gap as U goes from 2.65 to 2.7, as expected for this value of $\beta$. Notably, analytic continuation of the raw CTQMC data in this regime becomes extremely challenging due to the high amount of noise when approaching the transition. 
We test this low measurement scenario for this well-explored model to show that the features of a low-temperature phase diagram can be rapidly mapped using this technique. Phase diagram exploration of model Hamiltonians at low temperature remain challenging with CTQMC due to long autocorrelation lengths coupled with a requirement of large amounts of CTQMC steps. Here we show that by denoising the Green's function, one can reproduce qualitative features of the phase diagram rapidly. Not only is the calculation stabilized when approaching the metallic spinodal of this first-order transition, but we also recover spectral functions showing the correct Mott transition behavior with only 5000 CTQMC steps. 

\subsection{Dynamical Cluster Approximation Test}
We also test a multi-orbital model Hamiltonian as a more challenging benchmark of our denoising protocol. The two-patch dynamical cluster approximation (DCA)~\cite{dcaMaier2005} of the 2D Hubbard model has been previously studied a minimal model of the normal state of cuprate superconductors~\cite{vbhubbardFerrero2009}. It is defined on a square lattice with nearest and next-nearest neighbor interactions as
\begin{equation}
    H = \sum_{\mathbf{k},\sigma} \epsilon_\mathbf{k} c_{\sigma \mathbf{k}}^\dagger c_{\sigma \mathbf{k}} + U\sum_i n_{i \downarrow} n_{i \uparrow}
\end{equation}
where $\epsilon_{\mathbf{k}} = -2t(cos(k_x) + cos(k_y)) - 4t'cos(k_x)cos(k_y)$ and $t$ and $t'$ are the nearest and next-nearest neighbor interactions respectively. Following the treatment in Ref. \cite{vbhubbardFerrero2009}, we set $U/t=10$, $t'/t=-0.3$ and $t=0.25$.

By partitioning the Brillouin Zone (BZ) into 2 equal area parts, an effective 2-orbital impurity model can be defined and solved using the DCA. We treat the model within the paramagnetic state, meaning spin dependence is dropped and up and down spin components have equal Green's functions and hybridizations. We do not impose this condition during the DMFT loop however, leaving each spin-orbital to have its own independent self-energy. The regions within the BZ are the nodal $P_+$ patch enclosing the central region centered on $\mathbf{k}=(0,0)$ and the anti-nodal $P_-$ patch as the complement containing $\mathbf{k}=(\pi,0)$ and $\mathbf{k}=(0,\pi)$. 

By keeping the area of each patch the same, the local interaction term is preserved exactly and the model can be mapped to a diagonal 2-orbital problem. We aim to reproduce the results in Ref.~\cite{vbhubbardFerrero2009}, which used a few million Monte Carlo sweeps per iteration, with significantly fewer measurements. Here we use only $5\times10^{5}$ measurements, a tenfold reduction relative to our own $5\times10^{6}$-measurement reference. We set $\beta=200$ as in the original treatment, which presents challenges due to long autocorrelation times at low temperatures in the vicinity of the pseudogap transition. The ground state of the model was studied as a function of doping $\delta_c$, which is equivalent to adjusting the chemical potential, where it was found that the anti-nodal patch becomes insulating below $\delta_c\approx0.16$~\cite{vbhubbardFerrero2009}. We perform 10 DMFT self-consistent iterations for each doping value with $N_\mathrm{moves}=100$. 

\begin{figure*}[!t]
    \centering
    \subfloat[Denoised Result]{
        \includegraphics[width=0.82\columnwidth]{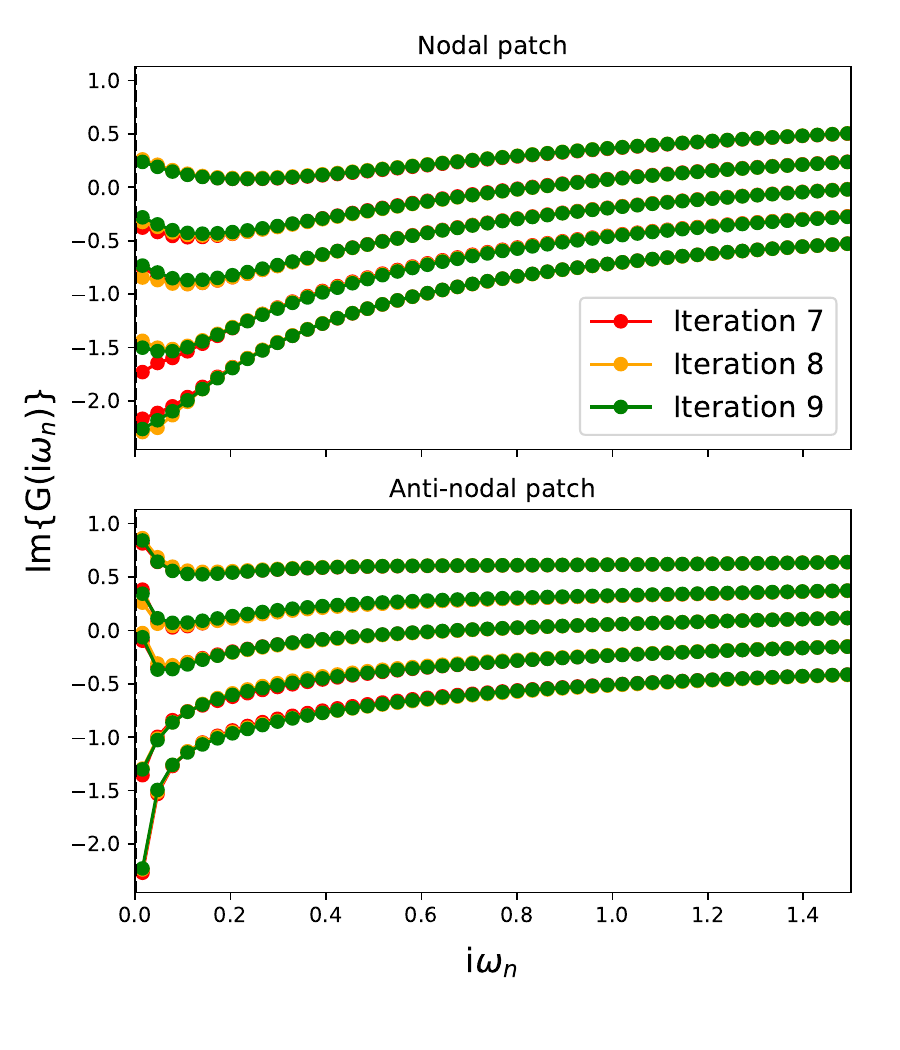}
        \label{subfig:denoisegiw}
    }
    \vspace{0.1cm}
    \subfloat[Raw CTQMC result]{
        \includegraphics[width=0.82\columnwidth]{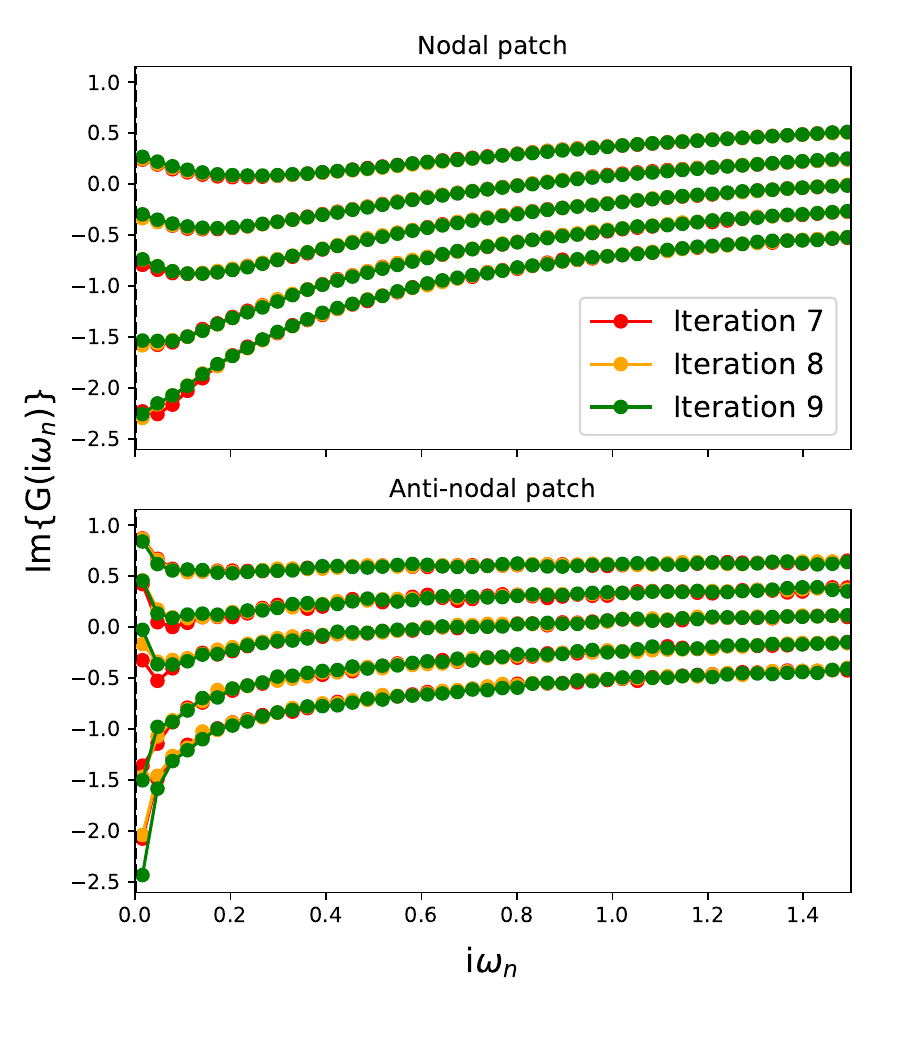}
        \label{subfig:rawgiw}
    }
    \caption{Imaginary part of $G(i\omega_n)$ for raw CTQMC and denoised CTQMC per DMFT iteration for various doping levels. Each doping level is offset by 0.25 for clarity. }
    \label{fig:vbhubdenoiseconverge}
\end{figure*}

In order to gauge the stability of the final solution, we plot the imaginary portion of the impurity Green's function for both the nodal and antinodal patches in Figure \ref{fig:vbhubdenoiseconverge} for the last 3 DMFT iterations. While both the denoised result and the raw CTQMC result display fluctuations consistent with Monte Carlo noise at this low sampling level, the raw CTQMC result is notably worse for the anti-nodal patch when undergoing the phase transition from metallic to insulating. The denoised result has fluctuations for certain doping levels in the nodal patch, but these fluctuations are minimal and can be attributed to different Monte Carlo seeds. 

\begin{figure*}[!t]
    \centering
    \subfloat[Denoised Result]{
        \includegraphics[width=0.82\columnwidth]{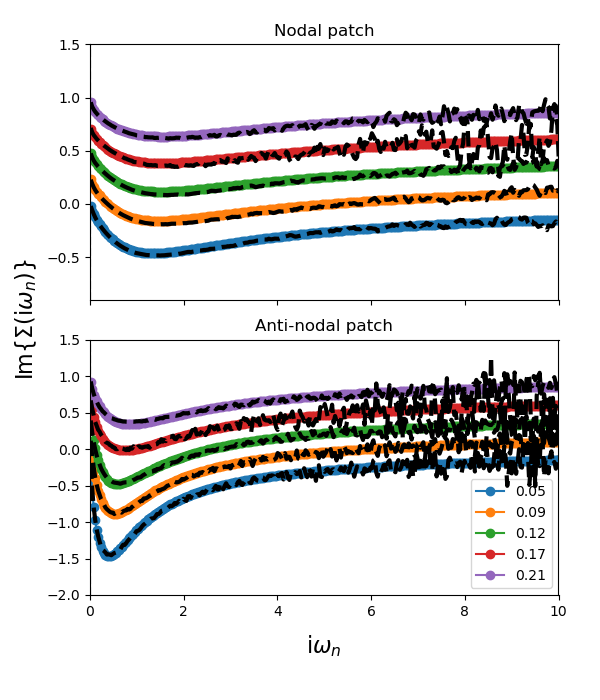}
        \label{subfig:denoisesiwcompare}
    }
    \vspace{0.1cm}
    \subfloat[Raw CTQMC result]{
        \includegraphics[width=0.82\columnwidth]{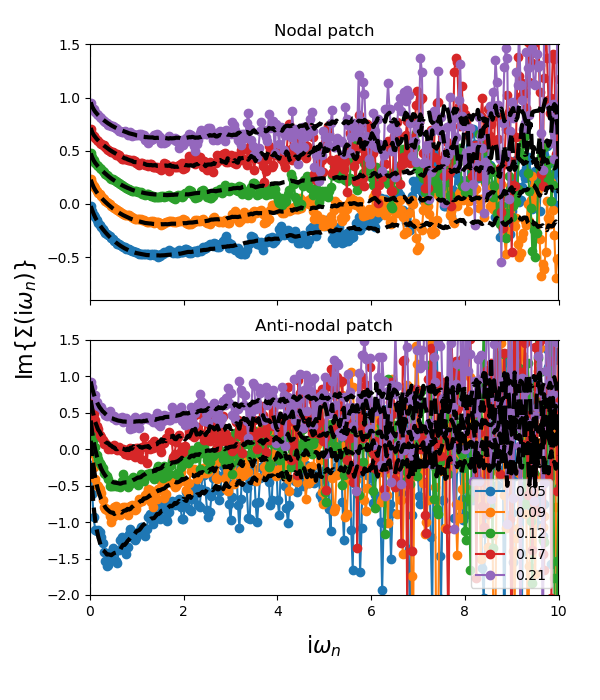}
        \label{subfig:lowrefsiwcompare}
    }
    \caption{Comparison of denoised versus raw CTQMC self energy against high step reference calculation at last DMFT iteration step. Reference calculation self-energy is shown by dashed black lines. }
    \label{fig:vbhubsiwcompare}
\end{figure*}

More noticeable is the noise in the predicted self-energies. As shown in Figure \ref{fig:vbhubsiwcompare}, the noise from the self energy grows enormous at intermediate Matsubara frequencies. Even the reference calculation gains significant noise, but may be remedied by some sort of tail fitting procedure. In contrast, the self-energy from the raw CTQMC calculations decoheres well before reaching the asymptotic tail behavior. Since our self-energy is optimized against more accurate moments through a CRM procedure, it displays the correct tail behavior automatically. The fact that it was generated from a positive normalized spectral function also means it is causal by construction~\cite{seproofStaar2014} without the instabilities that may arise when fitting DLR coefficients.

\begin{figure*}[!t]
    \centering
    \subfloat[Auxiliary spectral function]{
        \includegraphics[width=0.82\columnwidth]{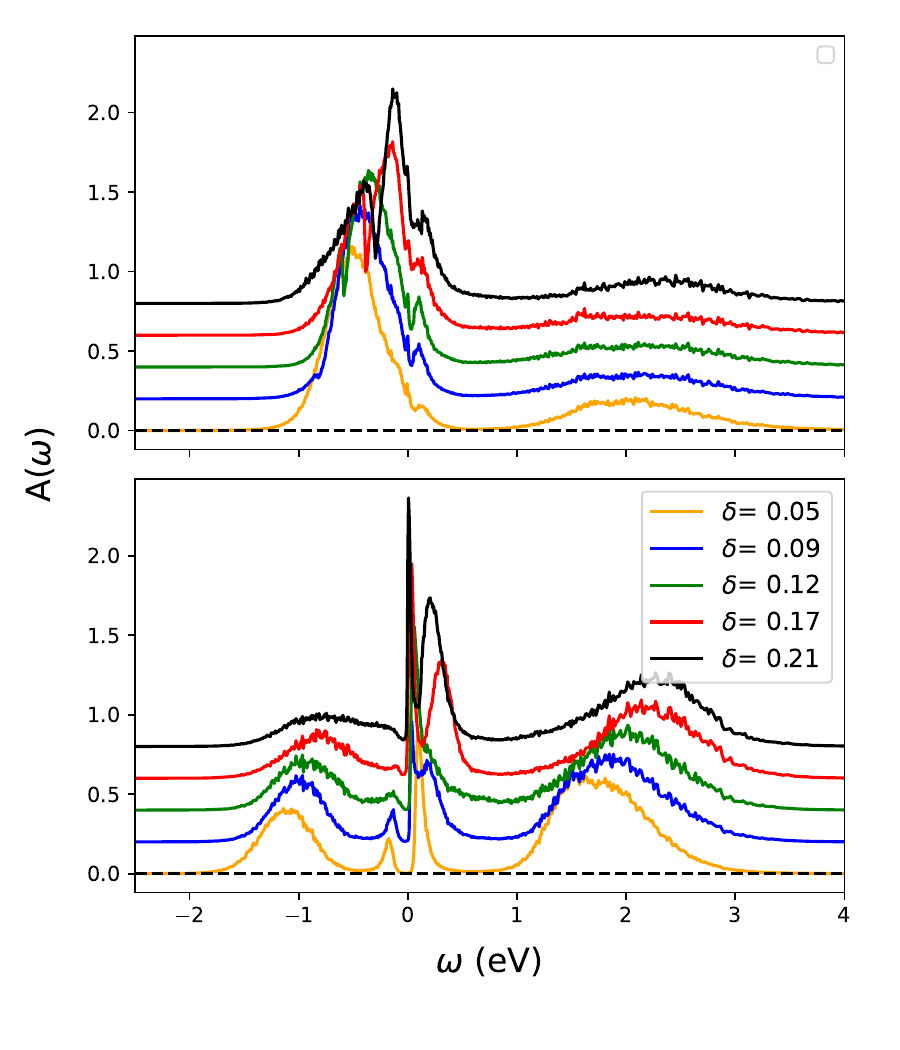}
        \label{fig:vbhubAw}
    }
    \subfloat[Pad\'e continuation of the denoised self-energy]{
        \includegraphics[width=0.82\columnwidth]{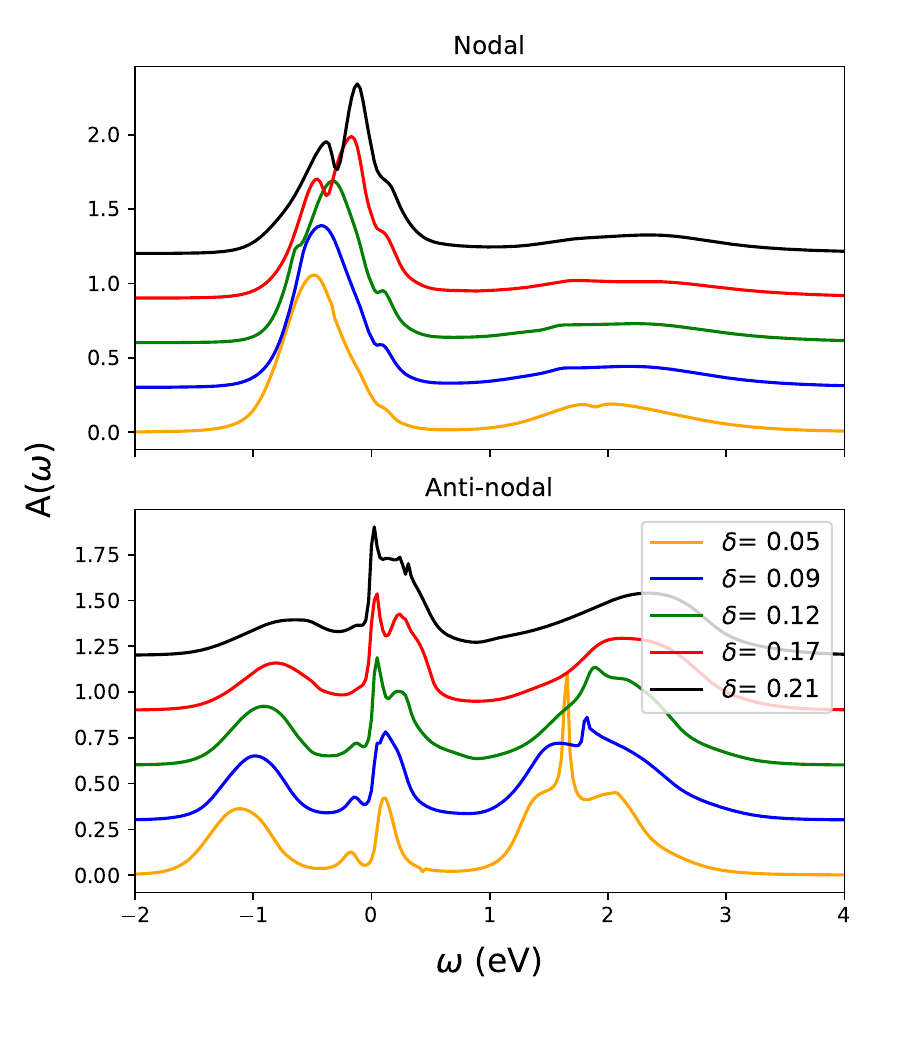}
        \label{fig:awfromsiwac_denoise}
    }
    \caption{Patch-resolved spectral functions of the two-patch DCA model as a function of doping (legend). (a) Internal spectral function generated by the denoising + optimization procedure. The anti-nodal patch transitions to a pseudogap phase as a function of doping. Data shifted upwards by 0.2 for clarity. (b) Spectral function obtained from analytic continuation of the denoised self-energy via Pad\'e approximants and integration over each patch. Data shifted upwards by 0.3 for clarity.}
    \label{fig:vbhubspectra}
\end{figure*}

Since we optimize an auxiliary spectral function towards the self-energy, we also obtain the spectral function for free through the calculation and ensure that its consistent with moment and density constraints. This provides a reference for gauging the quality of an analytic continuation since an auxiliary reference spectrum consistent with the measured data exists. Our auxiliary spectral function for the nodal and anti-nodal patches are shown in Figure \ref{fig:vbhubAw}. The spectra reproduce the findings in Ref \cite{vbhubbardFerrero2009} well, with the formation of a pseudogap at low energies as a function of doping. The general features away from the Fermi level also match as expected for the nodal and anti-nodal patches as a function of doping. Particularly, we see a transition of the lower Hubbard band towards the Fermi level, a transition away from the Fermi level for the upper Hubbard band and a spectral weight increase at the Fermi level as doping is increased.

Since we have perfectly smooth and causal self-energies, we also choose to perform an analytic continuation of the self-energy using Pad\'e approximants~\cite{padeVidberg1977} to recover the momentum integrated spectral function of each patch using
\begin{equation}
    A_\pm(\omega) = -\frac{1}{\pi} Im \sum_{k\in P_{\pm}} \frac{1}{\omega - (\epsilon_k - \mu) - \Sigma_{\pm}(\omega)}
\end{equation}
The results of this analytic continuation and integration for each doping level are shown in Figure \ref{fig:awfromsiwac_denoise}. General features seen in the auxiliary spectral function are maintained, with some smoothing of the curve due to Brillouin zone integration. We do see some unphysical features in the upper Hubbard band for doping levels of 0.05 and 0.09, but the qualitative features are generally preserved. 
Analytic continuation of the equivalently sampled noisy CTQMC self-energy is extremely challenging in this low step scenario, owing to the fact that noise takes over before tail behavior is reached. We attempted both Pade and maximum entropy analytic continuation of the noisy self-energy from CTQMC, but both methods failed to produce anything close to reasonable spectra.

\subsection{\protect\NoCaseChange{SrVO}$_3$}
\begin{figure*}[!t]
    \centering
    \subfloat[Self-energy]{
        \includegraphics[width=0.9\columnwidth]{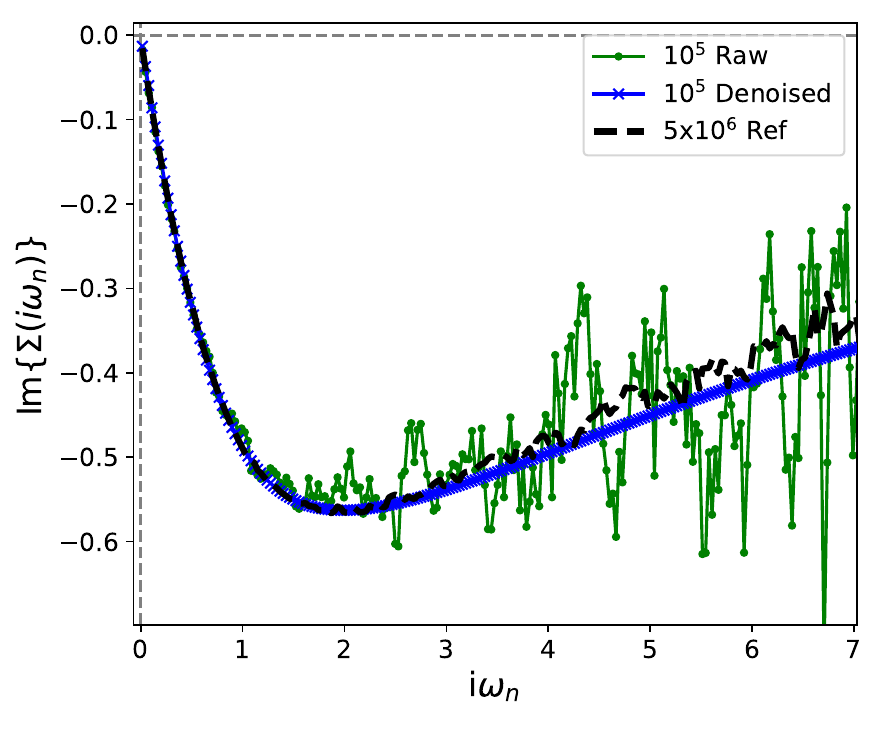}
        \label{fig:srvo3_siw_compare}
    }
    \subfloat[Internal spectral function]{
        \includegraphics[width=0.9\columnwidth]{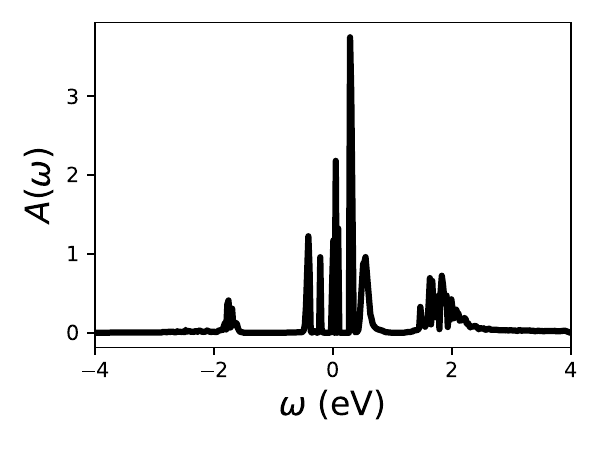}
        \label{fig:srvo3_internal_aw}
    }
    \caption{SrVO$_3$. (a) Imaginary part of the $t_{2g}$ self-energy for the denoised calculation with 10$^5$ CTQMC measurements versus the high ($5\times10^{6}$) and low ($10^{5}$) measurement reference cases. (b) Internal spectral function representation from the denoised calculation.}
    \label{fig:srvo3_siw_aw}
\end{figure*}

\begin{figure*}[!t]
    \centering
    \includegraphics[width=2.0\columnwidth]{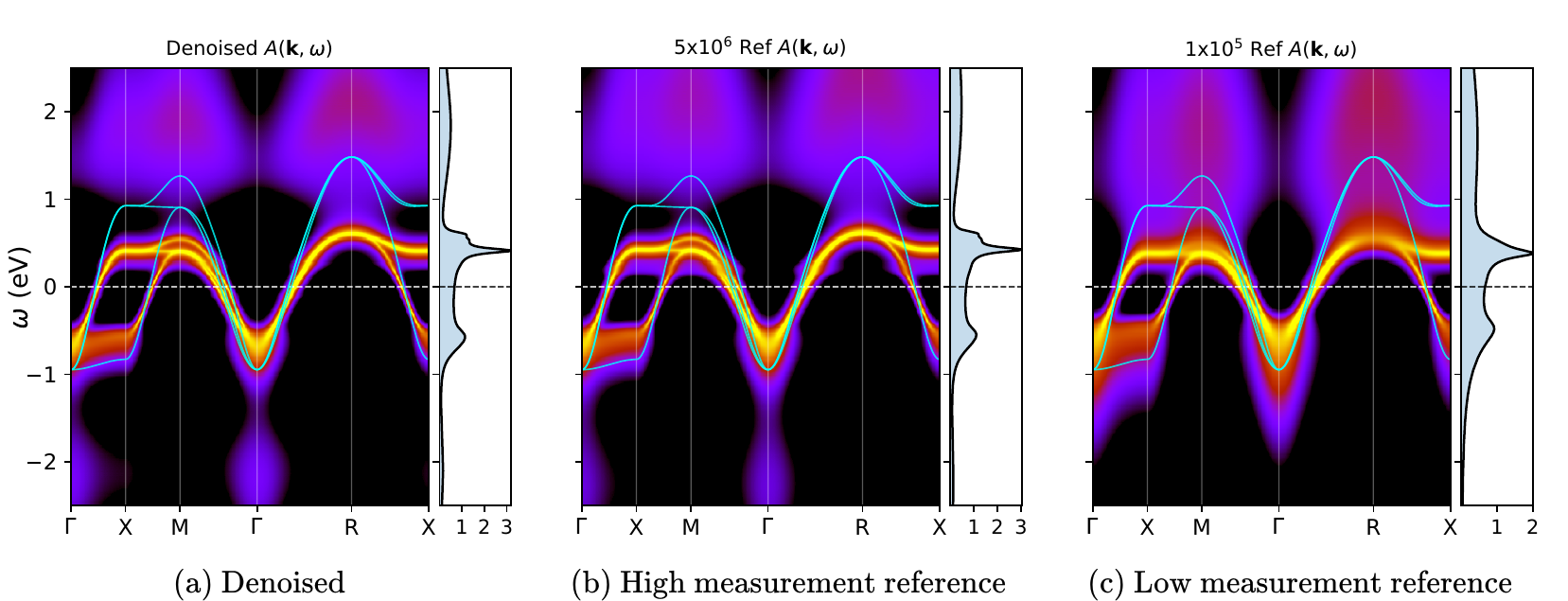}
    \caption{Comparison of denoised versus low and high measurement reference calculations of A(k, $\omega$)}
    \label{fig:srvo3_spec_func_compare}
\end{figure*}

Our final benchmark is SrVO$_3$, a correlated metal frequently used as a test system for DFT+DMFT~\cite{srvo3Pavarini2004,srvo3Sekiyama2004,srvo3Nekrasov2006,srvo3Aichhorn2009,srvo3Kugler2026}. SrVO$_3$ is a cubic perovskite with a 3d$^1$ configuration where the degenerate $t2g$ orbitals display strong band renormalization upon inclusion of correlation effects. Here we perform a DMFT self-consistent calculation of SrVO$_3$ starting from a Wannier Hamiltonian constructed from the $t2g$ orbitals of Vanadium. We utilize a Kanamori form of rotationally invariant Coulomb interaction with only density-density terms included so that the Green's function is diagonal. Coulomb interaction is set to $U=4$ and Hund's coupling to $J=0.6$. The value of $\beta$ is once again set to 200 eV$^{-1}$ to provide a challenging regime. We perform 15 DMFT iterations for 3 cases: a high measurement reference calculation with 5$\times$10$^6$ CTQMC measurements, a low measurement calculation with 10$^5$ measurements, and a denoised calculation with 10$^5$ measurements, using $N_\mathrm{moves}=200$ for each case.

Shown in Figure \ref{fig:srvo3_siw_compare} are the measured self-energies of the degenerate $t2g$ orbitals for each measurement scenario. The raw $10^5$ reference calculation shows significant noise before reaching the asymptotic regime, similar to our results for the DCA. The high-step reference calculation and the denoised calculation are in good agreement, especially at low frequencies. With this level of noise, analytic continuation of the low measurement calculation becomes quite challenging, while the denoised self-energy is perfectly smooth and can be analytically continued easily. The auxiliary spectral function representation is less indicative of the true spectral function here, but general features can still be seen. Shown in Figure \ref{fig:srvo3_internal_aw} is the internal spectral function representation. The lower and upper Hubbard bands are discernible, but the overall structure is composed of a series of sharp peaks. 

We also show the calculated momentum-dependent spectral functions $A(\mathbf{k},\omega)$ from analytic continuation of the self-energies in Figure \ref{fig:srvo3_spec_func_compare}. In all three cases, we have used the maximum entropy method~\cite{maxentJarrell1996} from the TRIQS/maxent package \cite{maxentKraberger2017,triqsmaxentgithub} to perform the analytic continuation. Since the denoised self-energy has zero noise, we add a very small amount of noise to stabilize the maximum entropy algorithm. The agreement between the spectral functions generated from analytically continuing the self-energy of the high measurement reference and the denoised calculations is excellent. In particular, the denoised and reference spectra agree in the position and width of the renormalized $t_{2g}$ quasiparticle bands, in the incoherent lower Hubbard band near $-1.5$~eV, in the upper Hubbard band near $2$~eV, and their integrated $A(\omega)$ agree quantitatively.

In contrast, the low measurement reference calculation shows significant departure from the high measurement reference. In particular, the upper and lower Hubbard bands are completely washed out and the resolution of sharp features is lost. The broadening of the coherent bands near the Fermi level is also indicative of the difficultly in obtaining a proper real-axis self energy. 
For the denoised and low measurement reference calculations, each DMFT iteration took 10 seconds on a 10 core CPU. In contrast, the high step reference calculation took around 8 minutes on 10 CPU cores while yielding similar results to the denoised calculation. This is a factor of 50$\times$ saving on CPU time to achieve comparable results. 

\section{Conclusion}
Here we have demonstrated a method of obtaining a massive reduction in required CTQMC steps to obtain a clean solution to the impurity problem within DMFT by utilizing a diffusion model based denoiser along with a CRM style optimization. By carrying an auxiliary spectral function representation, the causal nature of the predicted denoised imaginary axis quantities are preserved by construction. Our method combines the ideas of CRM style optimization with denoising diffusion models to obtain a joint framework that can operate well for any diagonal Green's function denoising task. Both model Hamiltonians and realistic systems were benchmarked, showing major performance improvements on the order of 10-100$\times$ reduction in CPU time to obtain a solution to the impurity problem. Our hope is that this method can be used to accelerate DFT+DMFT calculations for realistic systems in the future, lowering computational requirements and making them more accessible to researchers with lower computational budgets. Further studies integrating this method into the full DFT+DMFT charge self-consistency loop are planned and may provide a route to drastically lower the cost of achieving charge self-consistency. 

\textit{Acknowledgements}---This work was supported by the startup funds of the office of the Dean of SASN of Rutgers University-Newark. The authors acknowledge the Office of Advanced Research Computing (OARC) at Rutgers for providing access to the Amarel cluster and associated research computing resources.

\textit{Data Availability}---The denoiser, training data generation, and CRM-style refinement codes are available at \href{https://github.com/americandada29/GtauDenoise}{https://github.com/americandada29/GtauDenoise}

\bibliography{biblio}

\end{document}